\documentclass[aps,prb,twocolumn,longbibliography,floatfix]{revtex4-2}

\usepackage{graphicx}
\usepackage{url}
\usepackage{bm}
\usepackage{siunitx}
\usepackage{booktabs}
\usepackage{amsmath}
\usepackage[hidelinks]{hyperref}
\graphicspath{{./}}
\newcommand{\aSi}{5.476}
\newcommand{\aSiC}{4.390}
\newcommand{\BSi}{89}
\newcommand{\BSiC}{208}
\newcommand{\EgSi}{0.62}
\newcommand{\EgSiC}{1.34}

\newcommand{\dHfSiC}{-0.49}
\newcommand{\LSi}{10.95}

\newcommand{\EMPone}{0.16}
\newcommand{\EMPtwo}{0.64}

\newcommand{\EfVz}{3.46}

\newcommand{\nnminVHtwoz}{3.21}

\newcommand{\MagVHthreez}{0.91}

\newcommand{\EfVHfourz}{1.15}

\newcommand{\XHVHfourz}{1.486}

\newcommand{\EfVOz}{2.20}

\newcommand{\nnminVOz}{3.30}

\newcommand{\SiOVOz}{1.75}

\newcommand{\EfVOHz}{1.67}

\newcommand{\MagVOHz}{0.96}

\newcommand{\XHVOHz}{1.499}

\newcommand{\OHbVOHz}{2.095}

\newcommand{\EfVCzSirich}{3.90}

\newcommand{\EJTVCz}{0.19}
\newcommand{\nnminVCz}{2.80}

\newcommand{\XHVCHz}{1.651}
\newcommand{\EfVSizSirich}{7.64}

\newcommand{\dEspinVSiz}{0.07}

\newcommand{\XHVSiHz}{1.114}

\newcommand{\EfVCzCrich}{4.38}

\newcommand{\EfVSizCrich}{7.15}

\newcommand{\EbVH}{+0.44}
\newcommand{\EbVHtwo}{+0.60}
\newcommand{\EbVHthree}{+0.54}
\newcommand{\EbVHfour}{+0.73}
\newcommand{\EbVC}{-0.13}
\newcommand{\EbVSi}{+1.97}
\newcommand{\epsVPZ}{+0.01}
\newcommand{\epsVZM}{+0.78}
\newcommand{\epsVMMM}{+1.16}

\newcommand{\epsVUeff}{+0.76}

\newcommand{\epsVHfourPZ}{-0.21}
\newcommand{\epsVOZM}{+0.72}
\newcommand{\epsVOHZM}{+0.78}

\newcommand{\LVMVH}{2202}
\newcommand{\LVMexpVH}{2068}
\newcommand{\LVMmeanVH}{2202}

\newcommand{\LVMVHtwo}{2218, 2226}
\newcommand{\LVMexpVHtwo}{2121, 2144}
\newcommand{\LVMmeanVHtwo}{2222}

\newcommand{\LVMVHthree}{2185, 2223}
\newcommand{\LVMexpVHthree}{2166, 2191}
\newcommand{\LVMmeanVHthree}{2198}

\newcommand{\LVMVHfour}{2252 (2303 A$_1$, IR-inactive)}
\newcommand{\LVMexpVHfour}{2222}
\newcommand{\LVMmeanVHfour}{2253}
\newcommand{\LVMerrVHfour}{+1.4}
\newcommand{\LVMscale}{0.969}
\newcommand{\LVMmad}{69}
\newcommand{\LVMmadpct}{3.2}
\newcommand{\LVMmadscaled}{44}
\newcommand{\LVMmaxpct}{6.5}
\newcommand{\LVMminpct}{0.9}
\newcommand{\cutVO}{+0.30}
\newcommand{\cutVOH}{+0.26}

\newcommand{\cutmaxH}{0.04}

\newcommand{\dmuVtoVHfour}{-0.58}
\newcommand{\dmuVOHtoVHfour}{-0.10}
\newcommand{\dmuVOtoVOH}{-0.53}
\newcommand{\pHVHfour}{2}

\newcommand{\pHVOH}{9}
\newcommand{\Edisp}{0.23}
\newcommand{\EdispOdd}{0.29}

\newcommand{\EbO}{1.26}

\newcommand{\nnmeanVp}{3.82}
\newcommand{\nnmeanVmm}{3.33}
\newcommand{\nnideal}{3.87}
\newcommand{\dmuVtoVHfourG}{-0.34}
\newcommand{\pHVtoVHfourG}{6}
\newcommand{\dmuVOtoVOHG}{-0.29}
\newcommand{\pHVOHG}{5}
\newcommand{\dmuVOHtoVHfourG}{+0.02}

\newcommand{\UVHfourG}{-0.43}
\newcommand{\UVOHpHsevenG}{-0.12}
\newcommand{\ZPEperH}{0.21}

\newcommand{\ZPEO}{0.10}
\newcommand{\nuVOasym}{760}
\newcommand{\ZPEHtwo}{0.26}
\newcommand{\nuHtwo}{4222}
\newcommand{\ZPEHtwoO}{0.52}
\newcommand{\gHtwo}{-0.05}
\newcommand{\gHtwoO}{-0.05}
\newcommand{\shiftperH}{0.23}
\newcommand{\ZPESispread}{0.03}
\newcommand{\EdispG}{0.22}
\newcommand{\dmuVCH}{+0.13}
\newcommand{\dmuVSiH}{-1.97}

\newcommand{\VC}{$V_{\mathrm{C}}$}
\newcommand{\VSi}{$V_{\mathrm{Si}}$}
\newcommand{\EF}{E_F}
\newcommand{\dmuH}{\Delta\mu_{\mathrm{H}}}

\begin{document}

\title{Charge state and hydration of vacancies in silicon and silicon carbide in contact with water: A grand-canonical first-principles stability map}

\author{Abhirup Patra}
\email{patraa@udel.edu}
\affiliation{Department of Chemical \& Biomolecular Engineering, University of Delaware, Colburn Laboratory, 150 Academy Street, Newark, Delaware 19716, USA}

\begin{abstract}
The state of a vacancy in a semiconductor is normally described by the Fermi level alone. In wet-etched, electrochemically processed, or humid silicon and silicon carbide, a vacancy also takes up hydrogen and oxygen from water, in amounts set by pH and electrode potential. We treat both effects together by minimizing the grand potential of a vacancy complex over its charge state and hydrogen/oxygen content. The result is a stability map in Fermi level and hydrogen chemical potential, which the computational hydrogen electrode relates to pH and potential. We parameterize it with spin-polarized density functional theory in 64-atom cells, including the vibrational free energy of bound H and O, for the silicon vacancy, its H, O, and OH complexes, and both vacancies of 3C-SiC. Hydrogen binds cooperatively to the silicon vacancy, which goes directly from bare to fully passivated $V\mathrm{H}_4$. In contact with water the bare vacancy is never the lowest-energy state. Near neutral pH it binds oxygen from water as the A center, and full passivation to $V\mathrm{H}_4$ requires cathodic bias. In PBE the Fermi level barely moves these boundaries, because the acceptor levels lie at or near the conduction-band edge. In SiC the first hydrogen binds to the silicon vacancy by \SI{\EbVSi}{eV} but is unbound at the carbon vacancy (the $Z_{1/2}$ center), where it forms a weak three-center Si--H--Si bond.
\end{abstract}

\maketitle

\section{Introduction}

Point defects control carrier lifetime, compensation, and the optical and spin properties of semiconductors, and the first-principles treatment of their thermodynamics is well established~\cite{Freysoldt2014,VandeWalle2004,Kim2020}. The central quantity is the formation energy of a defect in a given charge state, which depends on the Fermi level $\EF$ and on the chemical potentials of the atoms exchanged with a reservoir. Formation energies and the charge-transition levels derived from them are used routinely to interpret deep-level transient spectroscopy (DLTS), electron paramagnetic resonance (EPR), and photoluminescence data~\cite{Freysoldt2014}.

The reservoir in most defect calculations is the growth atmosphere or a coexisting phase that fixes the stoichiometry. Many devices, however, spend much of their life in contact with water. Silicon and silicon carbide are wet-etched and photoelectrochemically processed~\cite{Kimoto2015}, they corrode in aqueous media, and they are operated at finite humidity. A vacancy in such a crystal exchanges electrons with the host and can at the same time pick up hydrogen and oxygen from the water. Hydrogen is the textbook amphoteric impurity: it passivates dangling bonds, forms a well-characterized family of vacancy--hydrogen complexes, and pins the Fermi level at a position tied to the vacuum scale~\cite{Pankove1991,VandeWalle1989,VandeWalle2003,Estreicher1995}. Oxygen forms the vacancy--oxygen A center, and hydroxyl groups introduce gap states of their own. In water the H and O supplies are linked by the dissociation equilibrium, and in an electrochemical cell both are also set by pH and electrode potential, so the electronic and chemical reservoirs have to be treated together.

Most of the ingredients for such a treatment exist. The computational hydrogen electrode (CHE) converts pH and potential into a hydrogen chemical potential and is the basis of first-principles surface analogues~\cite{Norskov2004,Hansen2008} of the Pourbaix diagram~\cite{Pourbaix1966}. For bulk solids, aqueous stability diagrams combine computed solid-state energies with experimental data for aqueous ions~\cite{Persson2012,Singh2017} and are now used in first-principles corrosion models~\cite{Huang2019}. Grand-canonical density functional theory treats the electrode potential at a solid--liquid interface explicitly~\cite{Sundararaman2017,Melander2019}. Todorova and Neugebauer showed that the defect-chemistry formalism of semiconductor physics carries over to electrochemical interfaces and used it to identify bulk oxide defects under electrochemical conditions~\cite{Todorova2014,Todorova2015}. We are not aware, however, of work that uses this coupling to follow a single bulk point defect from a bare, electrically active charge state to a decorated, passivated complex as the aqueous environment changes.

Here we write the formation energy of a vacancy complex as a grand potential and minimize it simultaneously over the charge state $q$ and the hydrogen/oxygen decoration $(m_{\mathrm{H}}, m_{\mathrm{O}})$, with $\mu_{\mathrm{H}}$ and $\mu_{\mathrm{O}}$ tied to a water reservoir through the CHE. The result is a two-dimensional stability map in $\EF$ and proton activity. We parameterize the model with density functional theory (DFT) for silicon and 3C-SiC in 64-atom supercells, treating spin explicitly and including the charged states of the hydrogenated complexes. After benchmarking the host crystals, the bare vacancy, and the Si--H local vibrational modes against experiment, we follow the hydrogenation, hydroxylation, and oxidation of the silicon vacancy on one diagram and then compare the two sublattice vacancies of SiC. The comparison shows that the first hydrogen binds strongly to the silicon vacancy in SiC but not to the carbon vacancy.

\section{Model}
\label{sec:model}

For a vacancy decorated with $m_{\mathrm{H}}$ hydrogen and $m_{\mathrm{O}}$ oxygen atoms in charge state $q$, the formation energy is the grand potential with respect to the electron, host, hydrogen, and oxygen reservoirs,
\begin{align}
\Omega[X^{q}](\EF,\mu_{\mathrm{H}},\mu_{\mathrm{O}})
 =\; & E_{\mathrm{tot}}[X^{q}] - E_{\mathrm{tot}}[\mathrm{host}] + \textstyle\sum_{s} n_s \mu_s \nonumber\\
 & - m_{\mathrm{H}}\mu_{\mathrm{H}} - m_{\mathrm{O}}\mu_{\mathrm{O}} \nonumber\\
 & + q\,(E_{\mathrm{VBM}}+\EF) + E_{\mathrm{corr}}(q).
\label{eq:omega}
\end{align}
Here $E_{\mathrm{tot}}[X^{q}]$ and $E_{\mathrm{tot}}[\mathrm{host}]$ are the total energies of the defective and pristine supercells, $n_s$ is the number of host atoms of species $s$ removed and $\mu_s$ their chemical potential, $E_{\mathrm{VBM}}$ is the host valence-band maximum, $\EF$ is measured from the VBM, and $E_{\mathrm{corr}}(q)$ is the electrostatic finite-size correction. We write $E^{f}$ for $\Omega$ at $\EF=0$ and $\dmuH=0$. At each point in reservoir space we expect to observe the complex with the lowest $\Omega$,
\begin{equation}
\Omega^\star(\EF,\mu_{\mathrm{H}},\mu_{\mathrm{O}}) = \min_{q,\,m_{\mathrm{H}},\,m_{\mathrm{O}}} \Omega[X^{q}].
\label{eq:min}
\end{equation}

Both $\mu_{\mathrm{H}}$ and $\mu_{\mathrm{O}}$ are fixed here by equilibrium with water. Writing $\mu_{\mathrm{H}}=\tfrac12 E[\mathrm{H_2}]+\dmuH$, the shift $\dmuH$ follows from the proton activity through the CHE~\cite{Norskov2004},
\begin{equation}
\dmuH(\mathrm{pH},U,T) = -eU - 2.303\,k_B T\,\mathrm{pH},
\label{eq:che}
\end{equation}
where $U$ is the electrode potential versus the standard hydrogen electrode (SHE). Oxygen then follows from the water molecule,
\begin{equation}
\mu_{\mathrm{O}} = \mu_{\mathrm{H_2O}} - 2\mu_{\mathrm{H}}, \qquad
\mu_{\mathrm{OH}} = \mu_{\mathrm{H_2O}} - \mu_{\mathrm{H}},
\label{eq:water}
\end{equation}
with $\mu_{\mathrm{H_2O}}$ set by the temperature and water activity. Using water instead of $\mathrm{O_2}$ as the oxygen reference avoids the known overbinding of the oxygen molecule in semilocal functionals, and it makes the H-rich (reducing) limit automatically O-poor. We treat $\EF$ and $\dmuH$ as independent variables. This is the appropriate limit when the bulk Fermi level is fixed by doping and the electrode potential enters only through the interfacial $\mathrm{H^+}/e^-$ couple that sets $\mu_{\mathrm{H}}$, i.e., when a depletion layer separates the bulk from the electrolyte. In full electronic equilibrium the two are not independent: $\EF$ is then fixed by $U$. Placing the band edges on the SHE scale, which lies \SI{4.44}{eV} below the vacuum level~\cite{Trasatti1986}, with the experimental electron affinities of Si (\SI{4.05}{eV}) and 3C-SiC (about \SI{4.0}{eV}) puts the valence-band maximum at about \SI{+0.73}{V} vs SHE for silicon and near \SI{+1.9}{V} for 3C-SiC. At $U=0$ a silicon crystal in equilibrium with the electrode would therefore have $\EF\approx E_v+\SI{0.73}{eV}$, above the PBE conduction-band minimum, and the pH guide lines in our maps refer to the decoupled case.

The model describes a constrained equilibrium. The vacancy exchanges electrons, hydrogen, and oxygen with the reservoirs, but the host lattice is held fixed: bulk silicon is unstable against oxidation to $\mathrm{SiO_2}$ over the whole range of conditions considered, and that reaction is not part of the model. The map therefore gives the lowest-energy decoration a vacancy can reach; whether it is reached depends on how fast hydrogen, hydroxyl, and the vacancy itself move. The oxygen supply in this regime is large. From the experimental free energy of $\mathrm{Si}+2\,\mathrm{H_2O}\to\mathrm{SiO_2}+2\,\mathrm{H_2}$, water at $\dmuH=0$ fixes $\mu_{\mathrm{O}}$ about \SI{2.0}{eV} above its value at Si/$\mathrm{SiO_2}$ coexistence. A vacancy in equilibrium with a native oxide sees the lower value, and the oxygen-containing complexes are then correspondingly less stable.

\begin{figure}[t]
\centering
\includegraphics[width=\columnwidth]{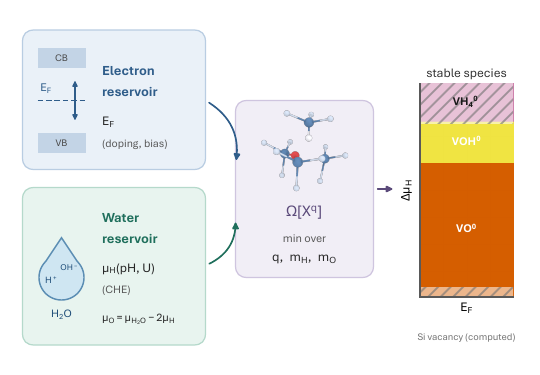}
\caption{Grand-canonical hydration model. The electron reservoir (Fermi level $\EF$, set by doping or bias) and the water reservoir, which fixes $\mu_{\mathrm{H}}$ through the computational hydrogen electrode and $\mu_{\mathrm{O}}$ through the water equilibrium, together determine the grand potential $\Omega[X^q]$ of each vacancy complex (shown: the relaxed $V$--OH complex). Minimizing over the charge state $q$ and the decoration $(m_{\mathrm{H}},m_{\mathrm{O}})$ gives a two-dimensional stability map; on the right, the computed map for the silicon vacancy, with hatched regions outside the stability window of water. Structure rendered with OVITO~\cite{Stukowski2010}.}
\label{fig:concept}
\end{figure}

\begin{figure*}[t]
\centering
\includegraphics[width=\textwidth]{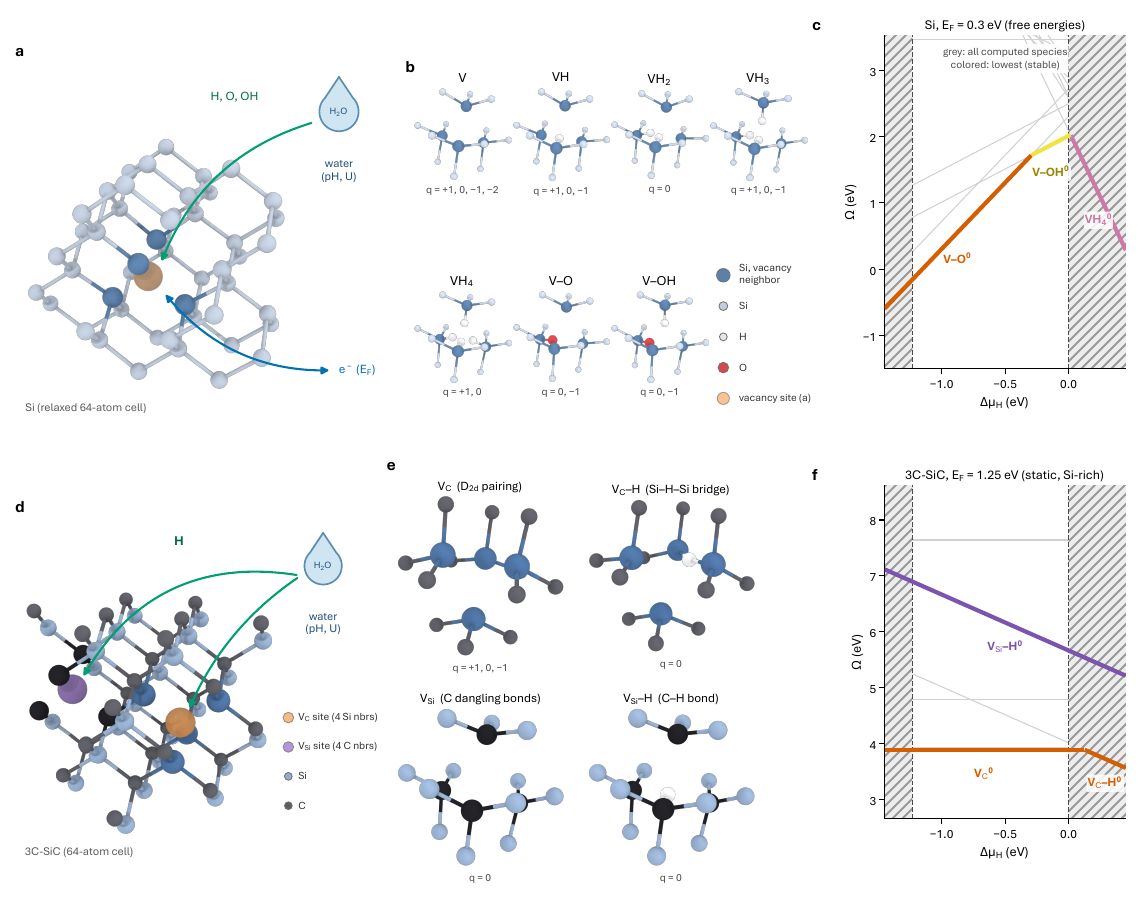}
\caption{Structures and stability of the vacancy complexes in silicon (a--c) and 3C-SiC (d--f). (a) Relaxed 64-atom silicon cell with a vacancy (orange; its four Si neighbors dark). Water is not simulated explicitly: it enters as a reservoir (drop icon) that exchanges H, O, and OH with the vacancy through $\mu_{\mathrm{H}}$ and $\mu_{\mathrm{O}}$, while electrons are exchanged with the reservoir at $\EF$. (b) Relaxed structures of the complexes computed for silicon: the four neighbors of the vacancy (dark) with their hydrogen and oxygen atoms and the next shell of silicon (light), with the charge states computed for each. (c) The grand potential $\Omega$ of every computed silicon species (grey) versus $\dmuH$ at $\EF=\SI{0.3}{eV}$, including vibrational free energies; the lowest line (colored) gives the stable species. (d) The 64-atom 3C-SiC cell with the carbon-vacancy site (orange, four Si neighbors) and the silicon-vacancy site (purple, four C neighbors) marked; the two vacancies were computed in separate cells, and for SiC only hydrogen exchange with the water reservoir was computed (arrows). (e) Relaxed structures of \VC{} ($D_{2d}$ pairing), \VC{}--H (hydrogen bridging two Si neighbors), \VSi{}, and \VSi{}--H (C--H bond), with the charge states computed. (f) $\Omega$ of the two SiC vacancies versus $\dmuH$ at $\EF=\SI{1.25}{eV}$, on the $n$-type side of the PBE gap where $V_{\mathrm{C}}^{0}$ is the stable bare state, at the Si-rich limit with static formation energies; the lower envelope of each vacancy is colored. In (c) and (f), hatched regions lie outside the stability window of liquid water. Structures rendered with OVITO~\cite{Stukowski2010}.}
\label{fig:structures}
\end{figure*}

Charged supercells interact spuriously with their periodic images. We apply the Makov--Payne monopole correction~\cite{Makov1995}
\begin{equation}
E_{\mathrm{MP}}(q) = \frac{q^2 \alpha e^2}{2\varepsilon L},
\label{eq:mp}
\end{equation}
where $\alpha=2.8373$ is the Madelung constant of the simple-cubic point-charge lattice, $\varepsilon$ the static dielectric constant, and $L$ the cell edge, and add a potential-alignment term $q\,\Delta V$. Here $\Delta V$ is the difference in the spherically averaged electrostatic potential between the defective and pristine cells at the atoms farthest from the defect~\cite{Lany2008,Komsa2012}. This is the leading term of the Freysoldt--Neugebauer--Van de Walle (FNV) scheme~\cite{Freysoldt2009}, and for a monovacancy in a 64-atom cubic cell of a high-dielectric host the two corrections agree closely~\cite{Komsa2012,Kumagai2014}. In our silicon cells $\Delta V$ is dominated by a composition term of \SIrange{-0.20}{-0.39}{eV} that is already present in the neutral cell; the charge-dependent part is at most \SI{0.13}{eV} ($V\mathrm{H}_4^{+}$). The composition term is physical in this setup: in the $V\mathrm{H}_4$ cell, which has no gap states, the valence- and conduction-band edges lie \SI{0.15}{} and \SI{0.11}{eV} below those of the pristine cell, consistent with its $\Delta V=\SI{-0.20}{eV}$. We therefore keep the full defect-minus-bulk alignment; the difference of \SIrange{0.04}{0.09}{eV} between $\Delta V$ and these band-edge shifts is a measure of its uncertainty. The thermodynamic charge-transition level of a given decoration is
\begin{equation}
\varepsilon(q/q') = \frac{E^{f}[X^{q};\EF{=}0] - E^{f}[X^{q'};\EF{=}0]}{q'-q}.
\label{eq:tl}
\end{equation}
Evaluating Eq.~(\ref{eq:min}) on a grid of $(\EF,\dmuH)$ and recording the lowest-energy $(q,m_{\mathrm{H}},m_{\mathrm{O}})$ at each point gives the defect stability map; Fig.~\ref{fig:concept} summarizes the workflow, and Fig.~\ref{fig:structures}c shows how the lower envelope of $\Omega$ selects the stable species.

\section{Methods}

Total energies were computed with the plane-wave projector augmented-wave (PAW) code GPAW~\cite{Enkovaara2010,Mortensen2005,Bloechl1994}, driven through the Atomic Simulation Environment~\cite{Larsen2017}, with the Perdew--Burke--Ernzerhof (PBE) exchange--correlation functional~\cite{Perdew1996}. We used the standard GPAW PBE PAW data sets for Si, C, O, and H and Fermi--Dirac smearing of \SI{0.05}{eV}; all solid-state calculations were spin-polarized. Geometries, charged-state energies, and potential alignments were obtained at a plane-wave cutoff of \SI{340}{eV}. The oxygen data set is far from converged in absolute energy at this cutoff, so every neutral formation and binding energy quoted in the text was recomputed as a single point at \SI{700}{eV} on the \SI{340}{eV} geometry, with the molecular references at the same cutoff. This changes the hydrogen-only complexes by at most \SI{\cutmaxH}{eV}, but $V$--O by \SI{\cutVO}{eV} and $V$--OH by \SI{\cutVOH}{eV}. The charged states of a decoration receive the same composition-dependent shift, so the transition levels were taken directly from the \SI{340}{eV} calculations.

All defects were modeled in a 64-atom cubic supercell, a $2\times2\times2$ repetition of the conventional cell, at the PBE equilibrium lattice constant of the host. The lattice constants, \SI{\aSi}{\angstrom} for Si and \SI{\aSiC}{\angstrom} for 3C-SiC, come from Birch--Murnaghan fits~\cite{Birch1947} with a \SI{400}{eV} cutoff and an $8\times8\times8$ mesh. The supercell Brillouin zone was sampled with a shifted $2\times2\times2$ Monkhorst--Pack mesh~\cite{Monkhorst1976}, which is coarser than the sampling found necessary for converged forces and formation energies of the silicon vacancy~\cite{Probert2003}; we discuss the consequences below. The host valence-band maximum and band gap were taken from a separate calculation of the pristine cell on a $\Gamma$-centered $4\times4\times4$ mesh, which samples the band edge at $\Gamma$ explicitly. Because the cell is the same and the densities are nearly identical, the two pristine calculations share the same potential reference.

Vacancies were created by removing one host atom (Si in silicon; Si or C in SiC, giving \VSi{} or \VC). Hydrogen atoms were placed \SI{1.50}{\angstrom} (Si--H) or \SI{1.10}{\angstrom} (C--H) from the nearest neighbors of the vacancy, along the neighbor-to-vacancy direction. In the $V$--O and $V$--OH complexes the oxygen started at the midpoint between two neighbors, the bridging geometry of the silicon A center, and the extra hydrogen of $V$--OH was put on a third neighbor. Starting geometries were pre-relaxed with the MACE-MP-0 universal interatomic potential~\cite{Batatia2024} and then fully relaxed with DFT. A pre-relaxed structure can sit on a symmetric saddle point, so before the DFT relaxation every defect cell was given random displacements of \SI{0.05}{\angstrom} per atom plus a \SI{0.15}{\angstrom} displacement of two dangling-bond atoms toward each other, and no symmetry was imposed. The symmetric solutions were also converged and used to obtain the Jahn--Teller relaxation energies. Odd-electron complexes were started from a doublet seed on a dangling-bond atom; even-electron complexes were started from both antiferromagnetic and ferromagnetic seeds, and the lower-energy solution was kept. When an odd-electron cell relaxed to a spin-unpolarized state, we also converged a fixed-spin-moment calculation for comparison, with the total moment constrained to \SI{1}{\mu_B} and a sharper smearing of \SI{0.01}{eV}. All atoms were relaxed with BFGS until the largest force was below \SI{0.03}{eV/\angstrom}. Charged cells were neutralized with a uniform background and corrected as described in Section~\ref{sec:model}, using the experimental static dielectric constants $\varepsilon=11.7$ (Si) and $9.72$ (3C-SiC). In the \SI{\LSi}{\angstrom} silicon cell the Makov--Payne term is \SI{\EMPone}{eV} for $|q|=1$ and \SI{\EMPtwo}{eV} for $|q|=2$.

The hydrogen and oxygen chemical potentials are referenced to $\mathrm{H_2}$ and $\mathrm{H_2O}$ molecules computed with the same basis and cutoff in a \SI{10}{\angstrom} vacuum box, which gives $\mu_{\mathrm{H}}^{\mathrm{ref}}=\tfrac12 E[\mathrm{H_2}]$ and, from Eq.~(\ref{eq:water}), $\mu_{\mathrm{O}}^{\mathrm{ref}}=E[\mathrm{H_2O}]-E[\mathrm{H_2}]$. In silicon the host chemical potential is the energy per atom of the pristine cell. In SiC the Si and C chemical potentials are bounded in the usual way~\cite{Zhang1991} by the Si-rich limit ($\mu_{\mathrm{Si}}$ equal to bulk silicon) and the C-rich limit ($\mu_{\mathrm{C}}$ equal to diamond, computed with the same settings). The two limits differ by the PBE formation enthalpy of SiC, \SI{\dHfSiC}{eV} per formula unit, and we give SiC vacancy formation energies at both. Hydrogen binding energies are differences at fixed host composition and do not depend on the limit. The sequential binding energy of the $m$th hydrogen relative to $\tfrac12\mathrm{H_2}$ is $E_{\mathrm{bind}}(m)=E^{f}[V\mathrm{H}_{m-1}]-E^{f}[V\mathrm{H}_{m}]$, so a positive value means that the hydrogen is bound with respect to the molecular reservoir.

Si--H stretching frequencies were calculated from the harmonic dynamical matrix of the hydrogen atoms and their four silicon neighbors, built from central finite differences of the forces (\SI{\pm 0.015}{\angstrom}). For these calculations the relaxed 64-atom complexes were relaxed further at the $\Gamma$ point to \SI{0.01}{eV/\angstrom}. We diagonalized the mass-weighted Hessian and took as stretch modes those above \SI{1500}{cm^{-1}} with more than \SI{50}{\percent} hydrogen participation. Frequencies in the text are unscaled. For the $T_d$ complex $V\mathrm{H}_4$ only the infrared-active $T_2$ triplet is compared with experiment; the Raman-only $A_1$ mode is given in parentheses.

The same Hessians, extended to the oxygen atom for $V$--O and $V$--OH, give the vibrational free energy of the bound hydrogen and oxygen. In every partial Hessian the twelve lowest modes are dominated by the four silicon neighbors and are common to all complexes (their zero-point energies differ by at most \SI{\ZPESispread}{eV}, which limits the accuracy of this treatment to about \SI{0.01}{eV} per H); the remaining three modes per H or O atom are local, and only these enter the free energy. The hydrogen modes (one stretch and two bends per H) carry a zero-point energy of \SI{\ZPEperH}{eV} per bound H in every $V\mathrm{H}_m$ complex. The oxygen modes of $V$--O, with the Si--O--Si asymmetric stretch at \SI{\nuVOasym}{cm^{-1}}, carry \SI{\ZPEO}{eV}. For the references we relaxed $\mathrm{H_2}$ and $\mathrm{H_2O}$ and computed their harmonic frequencies ($\mathrm{H_2}$: \SI{\nuHtwo}{cm^{-1}}, zero-point energy \SI{\ZPEHtwo}{eV}; $\mathrm{H_2O}$: \SI{\ZPEHtwoO}{eV}), then added the standard enthalpy and entropy at \SI{298}{K}, representing liquid water by the vapor at \SI{0.035}{bar}~\cite{Norskov2004}. The resulting free-energy corrections are \SI{\gHtwo}{eV} for $\mathrm{H_2}$ and \SI{\gHtwoO}{eV} for $\mathrm{H_2O}$. The oxygen-containing modes were computed at \SI{340}{eV}, where the O--H stretches of $\mathrm{H_2O}$ come out about \SI{10}{\percent} low; this underestimates the $\mathrm{H_2O}$ correction by about \SI{0.04}{eV} and moves the $V$--OH/$V\mathrm{H}_4$ boundary by about \SI{0.01}{eV}. Relative to the static result, each bound H raises the formation energy of a complex by \SI{\shiftperH}{eV}. Charged states take the vibrational term of the neutral complex with the same decoration. The stability maps and all pH and potential values below include these terms, and $\dmuH$ is measured from $\tfrac12 G(\mathrm{H_2})$ at \SI{298}{K} and \SI{1}{bar}, so that Eq.~(\ref{eq:che}) applies directly. Formation and binding energies quoted in the text are static unless stated otherwise.

\section{Results}

\begin{figure*}[t]
\centering
\includegraphics[width=\textwidth]{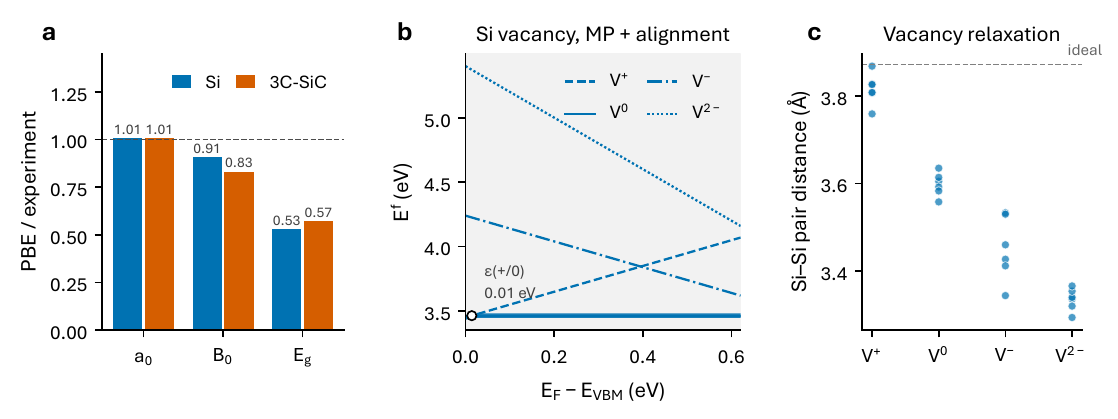}
\caption{Validation. (a) PBE bulk properties of Si and 3C-SiC normalized to experiment: lattice constant $a_0$, bulk modulus $B_0$, and band gap $E_g$. (b) Corrected formation energy of the silicon monovacancy in charge states $+1$ to $-2$ versus Fermi level (Makov--Payne with potential alignment, 64-atom cell). The thick line is the lower envelope; in PBE it is the neutral state across the whole gap. (c) Distances between the neighbors of the relaxed vacancy in each charge state. The dashed line is the ideal-lattice value.}
\label{fig:valid}
\end{figure*}

Figure~\ref{fig:valid}a compares the two hosts with experiment. The PBE lattice constants are within \SI{1}{\percent} of the measured values (\SI{\aSi}{} vs \SI{5.431}{\angstrom} for Si; \SI{\aSiC}{} vs \SI{4.360}{\angstrom} for 3C-SiC), the bulk moduli are \SIrange{10}{20}{\percent} too small (\SI{\BSi}{} vs \SI{97.8}{GPa}; \SI{\BSiC}{} vs \SI{\sim250}{GPa}), and, as usual for a semilocal functional, the Kohn--Sham gaps are too small (\SI{\EgSi}{} vs \SI{1.17}{eV}; \SI{\EgSiC}{} vs \SI{2.36}{eV}). The gap error matters most for this work. Transition levels are placed inside a gap only about half its true width, so their absolute positions are semiquantitative; relative quantities such as binding energies, reaction energies between decorations, and vibrational frequencies are much less affected~\cite{Deak2010}.

In the 64-atom cell the neutral silicon vacancy has a formation energy of \SI{\EfVz}{eV}, about \SI{0.3}{eV} above the \SI{3.17}{eV} obtained with the PW91 functional in a converged 255-atom cell~\cite{Probert2003}; the slow convergence of this quantity with cell size is analyzed in Ref.~\cite{Corsetti2011}. Figure~\ref{fig:valid}b shows the corrected formation energies of the $+1$ to $-2$ charge states as a function of Fermi level. In PBE the neutral state is lowest across the whole (underestimated) gap. The donor level $\varepsilon(+/0)$ is at the valence-band edge (\SI{\epsVPZ}{eV}), $\varepsilon(0/-)$ at \SI{\epsVZM}{eV}, i.e., \SI{0.16}{eV} above the PBE conduction-band minimum, and $\varepsilon(-/2-)$ at \SI{\epsVMMM}{eV}. The effective correlation energy is therefore positive, $U_{\mathrm{eff}}=\SI{\epsVUeff}{eV}$. Experimentally the silicon vacancy is an Anderson negative-$U$ center~\cite{Anderson1975}, with donor levels $\varepsilon(0/+)\approx E_v+\SI{0.05}{eV}$ and $\varepsilon(+/2+)\approx E_v+\SI{0.13}{eV}$~\cite{Watkins1980}; the negative $U$ comes from the strong Jahn--Teller relaxation of the charged states. We find the donor level at the valence-band edge but do not recover the negative-$U$ ordering; $V^{2+}$ was not computed, so only $\varepsilon(+/0)$ and $\varepsilon(0/-)$ can be compared. With \SI{0.05}{eV} smearing the seeded neutral vacancy ends \SI{1}{meV} from the symmetric solution, with neighbor distances of \SIrange{3.56}{3.63}{\angstrom} against \SI{3.61}{\angstrom}, so the pairing mode is soft but not unstable. The charged states contract inward, from a mean neighbor distance of \SI{\nnmeanVp}{\angstrom} for $V^{+}$ to \SI{\nnmeanVmm}{\angstrom} for $V^{2-}$ (ideal value \SI{\nnideal}{\angstrom}), with spreads of up to \SI{0.19}{\angstrom} ($V^{-}$) but no clear pairing (Fig.~\ref{fig:valid}c). For $V^{\pm}$ a fixed-moment calculation with \SI{0.01}{eV} smearing gives the same geometries, but $V^{2-}$ carries a fractional moment of \SI{1.38}{\mu_B}, a sign that the smearing spreads the occupation of the degenerate $t_2$ level. To test whether the smearing suppresses the distortion, we re-relaxed $V^{0}$ with \SI{0.01}{eV} smearing from a $D_{2d}$ seed in which both dangling-bond pairs were moved \SI{0.20}{\angstrom} toward each other. It returned to the same nearly symmetric structure (neighbor distances \SIrange{3.56}{3.62}{\angstrom}), within \SI{3}{meV} of the \SI{0.05}{eV} result. The smearing is therefore not the cause. With converged sampling in a 255-atom cell and the closely related PW91 functional, $V^{0}$ relaxes spontaneously to $D_{2d}$~\cite{Probert2003}, which points to our $2\times2\times2$ mesh and 64-atom cell rather than to the functional. We therefore use the bare-vacancy charge states only to place the charged species on the stability map and draw no conclusions from the absolute level positions.

\begin{figure*}[t]
\centering
\includegraphics[width=\textwidth]{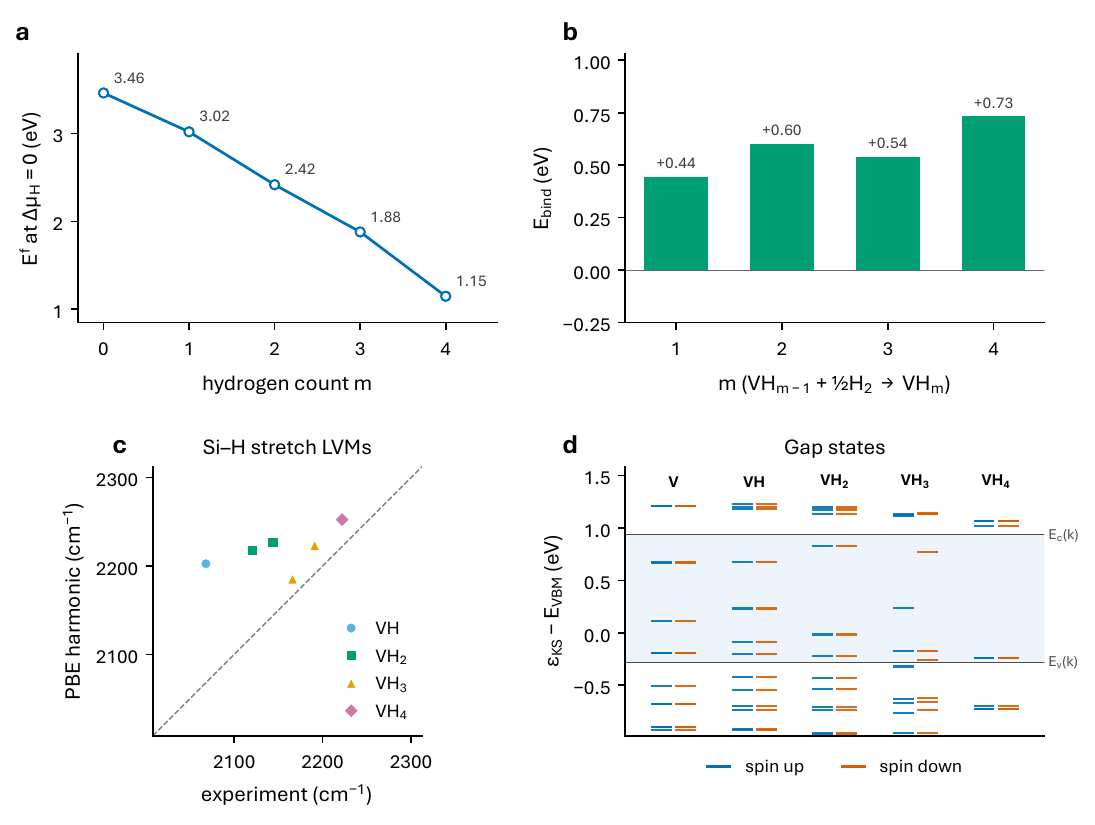}
\caption{Hydrogenation of the silicon vacancy. (a) Formation energy of the neutral $V\mathrm{H}_m$ complexes at $\dmuH=0$. (b) Sequential binding energy of each added hydrogen. (c) Calculated infrared-active Si--H stretch frequencies of $V\mathrm{H}_{1\text{--}4}$ (harmonic, unscaled) versus the infrared values of Ref.~\cite{BechNielsen1996}; the dashed line marks perfect agreement. (d) Spin-resolved Kohn--Sham eigenvalues near the gap at one $k$-point of the $2\times2\times2$ mesh for the neutral complexes, aligned by $\Delta V$; the shaded band spans the pristine band edges at the same $k$-point.}
\label{fig:hydro}
\end{figure*}

Each added hydrogen (structures in Fig.~\ref{fig:structures}b) lowers the formation energy of the neutral vacancy, from \SI{\EfVz}{eV} for $V$ to \SI{\EfVHfourz}{eV} for $V\mathrm{H}_4$ at the $\tfrac12\mathrm{H_2}$ reference (Fig.~\ref{fig:hydro}a). All four hydrogens are bound, with sequential binding energies of \SI{\EbVH}{}, \SI{\EbVHtwo}{}, \SI{\EbVHthree}{}, and \SI{\EbVHfour}{eV} for $m=1$--$4$ (Fig.~\ref{fig:hydro}b). On average each hydrogen binds more strongly than the one before it, so the hydrogenation is cooperative. We attribute this to strain: each new Si--H bond relieves strain on the remaining dangling bonds. In the intermediate complexes those dangling bonds pair up (the two unpassivated neighbors in $V\mathrm{H}_2$ move to \SI{\nnminVHtwoz}{\angstrom} of each other), whereas in $V\mathrm{H}_4$ they are replaced by four independent Si--H bonds of \SI{\XHVHfourz}{\angstrom}, close to the \SI{1.48}{\angstrom} bond in silane. The partially hydrogenated complexes are consequently unstable against disproportionation: $2\,V\mathrm{H}_2 \to V + V\mathrm{H}_4$ is exothermic by \SI{\Edisp}{eV}, and $V\mathrm{H}+V\mathrm{H}_3\to V+V\mathrm{H}_4$ by \SI{\EdispOdd}{eV}. The intermediate complexes that dominate the infrared spectra of hydrogen-implanted silicon~\cite{BechNielsen1996} are therefore kinetic products, formed when hydrogen is scarce, and do not appear on the equilibrium map below.

The gap states follow the number of Si--H bonds (Fig.~\ref{fig:hydro}d). The bare vacancy has a Kohn--Sham level near mid-gap. $V\mathrm{H}$ and $V\mathrm{H}_2$ still have gap states, which move toward the conduction band as the leftover dangling bonds pair. $V\mathrm{H}_3$ is a doublet (\SI{\MagVHthreez}{\mu_B}) whose singly occupied dangling-bond level lies \SI{0.5}{eV} below its empty spin partner, and $V\mathrm{H}_4$ has no gap state at all, i.e., it is fully passivated~\cite{VandeWalle2003,Estreicher1995}. The spin of $V\mathrm{H}$ does not localize in PBE; the constrained doublet is \SI{0.07}{eV} higher in energy than the unpolarized solution. We attribute this to the delocalization error of semilocal functionals for a single electron shared by three dangling bonds.

The Si--H local vibrational modes (LVMs) are the most direct experimental test of these structures. The infrared stretch modes lie at \SI{2068}{cm^{-1}} ($V\mathrm{H}$), \SI{2121}{} and \SI{2144}{cm^{-1}} ($V\mathrm{H}_2$), \SI{2166}{} and \SI{2191}{cm^{-1}} ($V\mathrm{H}_3$), and \SI{2222}{cm^{-1}} ($V\mathrm{H}_4$), and they rise with $m$ because of the H--H repulsion inside the vacancy~\cite{BechNielsen1996}. Our harmonic PBE frequencies (Table~\ref{tab:lvm}, Fig.~\ref{fig:hydro}c) fall in the right band but compress the spread between complexes: the mean IR-active stretch is \SI{\LVMmeanVH}{}, \SI{\LVMmeanVHtwo}{}, \SI{\LVMmeanVHthree}{}, and \SI{\LVMmeanVHfour}{cm^{-1}} for $m=1$--4, a range of about \SI{55}{cm^{-1}} against \SI{154}{cm^{-1}} in experiment, and $V\mathrm{H}_3$ comes out below $V\mathrm{H}_2$. Unscaled, the frequencies are \SI{\LVMminpct}{}--\SI{\LVMmaxpct}{\percent} above experiment (mean absolute deviation \SI{\LVMmad}{cm^{-1}}, or \SI{\LVMmadpct}{\percent}), with the largest overestimate for $V\mathrm{H}$ and the smallest for $V\mathrm{H}_3$. A single least-squares scale factor of \LVMscale{}, typical of harmonic PBE X--H stretches, reduces the mean deviation to \SI{\LVMmadscaled}{cm^{-1}}. The $V\mathrm{H}_4$ mode is also sensitive to cell size. In our preliminary 16-atom calculations it came out \SI{11}{\percent} too high; in the 64-atom cell, where the four Si--H bonds are no longer compressed by their periodic images, the infrared-active $T_2$ mode is within \SI{\LVMerrVHfour}{\percent} of the measured line. The frequencies support the computed $V\mathrm{H}_4$ geometry, but at this level of accuracy they do not discriminate between the intermediate complexes.

\begin{table}[t]
\caption{Si--H stretch local vibrational modes (cm$^{-1}$) of the vacancy--hydrogen complexes in silicon: harmonic PBE (this work, unscaled, 64-atom cell) and infrared experiment~\cite{BechNielsen1996}. Degenerate modes are listed once.}
\label{tab:lvm}
\begin{tabular}{lcc}
\toprule
Complex & PBE & Experiment\\
\midrule
$V\mathrm{H}$   & \LVMVH      & \LVMexpVH\\
$V\mathrm{H}_2$ & \LVMVHtwo   & \LVMexpVHtwo\\
$V\mathrm{H}_3$ & \LVMVHthree & \LVMexpVHthree\\
$V\mathrm{H}_4$ & \LVMVHfour  & \LVMexpVHfour\\
\bottomrule
\end{tabular}
\end{table}

Started from a bridging geometry, the vacancy--oxygen complex relaxes to the A-center structure (Fig.~\ref{fig:structures}b): the oxygen bridges two silicon neighbors (Si--O \SI{\SiOVOz}{\angstrom}) and pulls them to \SI{\nnminVOz}{\angstrom} of each other, the other two neighbors stay \SI{3.70}{\angstrom} apart, and the complex is closed-shell. In the $V$--OH complex the added hydrogen caps one of the two non-bridged silicon atoms instead of binding to the oxygen (Si--H \SI{\XHVOHz}{\angstrom}, O$\cdots$H \SI{\OHbVOHz}{\angstrom}). This leaves one dangling bond with a localized spin of \SI{\MagVOHz}{\mu_B}. At the water reference both $V$--O (\SI{\EfVOz}{eV}) and $V$--OH (\SI{\EfVOHz}{eV}) have formation energies far below that of the bare vacancy. Oxygen from water is therefore bound to the vacancy by \SI{\EbO}{eV} at $\dmuH=0$, and more strongly still under O-rich (H-poor) conditions.

\begin{figure*}[t]
\centering
\includegraphics[width=\textwidth]{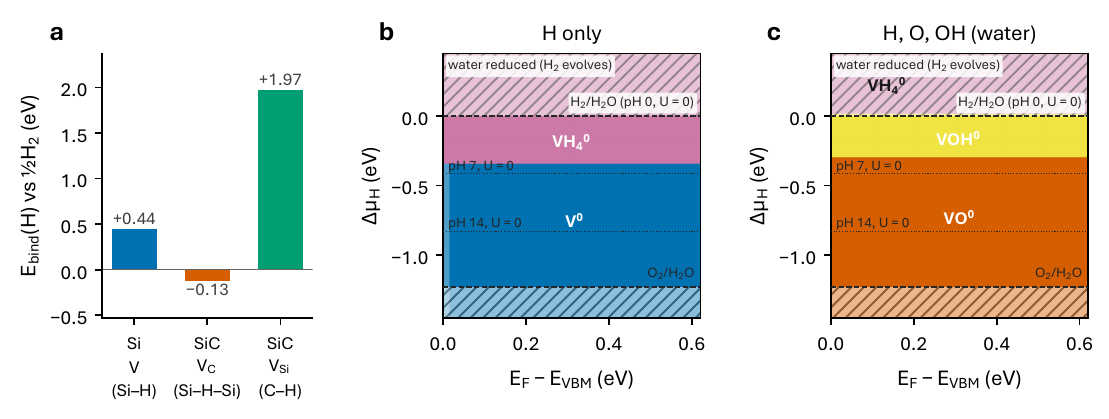}
\caption{(a) Static binding energy of the first hydrogen relative to $\tfrac12\mathrm{H_2}$ at the vacancy in silicon (Si--H) and at the carbon and silicon vacancies of 3C-SiC (Si--H and C--H). (b) Grand-canonical stability map of the silicon vacancy versus Fermi level and hydrogen chemical potential, including vibrational free energies ($\dmuH$ measured from $\tfrac12 G(\mathrm{H_2})$ at \SI{298}{K}, \SI{1}{bar}), with hydrogen only, including the charged states of $V$, $V\mathrm{H}$, and $V\mathrm{H}_3$. Regions are labeled by the stable species; the thin strip at the left edge is $V^{+}$. Dashed lines mark the $\mathrm{H_2}/\mathrm{H_2O}$ equilibrium ($\mathrm{H_2}$ at \SI{1}{bar}, equivalently pH~0 at $U=0$) and the $\mathrm{O_2}/\mathrm{H_2O}$ equilibrium; hatched regions lie outside the stability window of liquid water. Dotted lines mark pH 7 and 14 at $U=\SI{0}{V}$ vs SHE for a Fermi level decoupled from the electrode. (c) The same map with the $V$--O and $V$--OH complexes included; oxygen is referenced to liquid water through Eq.~(\ref{eq:water}). Here $V\mathrm{H}_4$ is stable only above the $\mathrm{H_2}$ line, and the oxygen complexes considered are limited to $V$--O and $V$--OH.}
\label{fig:model}
\end{figure*}

Figure~\ref{fig:model}b,c shows the minimum of Eq.~(\ref{eq:min}) over all the species we computed: the bare vacancy in four charge states, $V\mathrm{H}$ and $V\mathrm{H}_3$ in three charge states each, neutral $V\mathrm{H}_2$ and $V\mathrm{H}_4$, and the oxygen-containing complexes. The vertical axis is the shift $\dmuH$ of the hydrogen chemical potential from the $\tfrac12\mathrm{H_2}$ reference. The dashed line marks equilibrium with $\mathrm{H_2}$ gas at \SI{1}{bar}, and the dotted lines mark pH 0, 7, and 14 at zero bias versus SHE [Eq.~(\ref{eq:che})]. A cathodic bias moves the system up the plot, an anodic bias down.

With hydrogen only (Fig.~\ref{fig:model}b) the map has two regions. The bare neutral vacancy is stable below $\dmuH=\SI{\dmuVtoVHfourG}{eV}$ (static references: \SI{\dmuVtoVHfour}{eV}), i.e., above about pH~\pHVtoVHfourG{} at $U=0$. Above this value the fully passivated $V\mathrm{H}_4$ takes over directly, with no intermediate $V\mathrm{H}_m$ region, as expected from the cooperative binding; disproportionation of $V\mathrm{H}_2$ remains exothermic by \SI{\EdispG}{eV} with the vibrational terms. The thin strip at $\EF<\SI{0.01}{eV}$ belongs to $V^{+}$. No other charged species appear, because in PBE the acceptor levels of $V$, $V\mathrm{H}$, and $V\mathrm{H}_3$ lie at or just above the conduction-band edge (for $V\mathrm{H}$ by only \SI{0.03}{eV}, less than the uncertainty of the alignment) and their donor levels at the valence-band edge. The other complexes behave the same way: the acceptor levels of $V$--O and $V$--OH lie at \SI{\epsVOZM}{} and \SI{\epsVOHZM}{eV}, above the PBE gap of \SI{\EgSi}{eV}, and the donor level of $V\mathrm{H}_4$ lies at \SI{\epsVHfourPZ}{eV}, below the valence-band edge. (The positive states of $V$--O and $V$--OH and the negative state of $V\mathrm{H}_4$ were not computed.) Within PBE the map is therefore essentially independent of $\EF$. With the experimental gap, $V^{-}$ and $V^{2-}$ would occupy the high-$\EF$ part of the bare-vacancy region, and its boundary with neutral $V\mathrm{H}_4$ would tilt. Such a tilt would be a direct signature of the coupling between the two reservoirs; in a purely chemical treatment at fixed charge, every boundary is horizontal.

The oxygen-containing complexes change the map considerably (Fig.~\ref{fig:model}c). Since oxygen is referenced to water, H-poor conditions are also O-rich, and the vacancy--oxygen complex replaces the bare vacancy over the whole accessible range. In contact with water the electrically active bare vacancy is thus never the thermodynamic ground state. From H-poor to H-rich conditions the stable species are $V$--O below $\dmuH=\SI{\dmuVOtoVOHG}{eV}$, $V$--OH between \SI{\dmuVOtoVOHG}{} and \SI{\dmuVOHtoVHfourG}{eV}, and $V\mathrm{H}_4$ above (static references: \SI{\dmuVOtoVOH}{} and \SI{\dmuVOHtoVHfour}{eV}, i.e., pH~\pHVOH{} and pH~\pHVHfour{} at zero bias). At $U=0$ the $V$--O/$V$--OH boundary therefore lies near pH~\pHVOHG{}; at pH~7 the hydroxyl complex needs a cathodic bias of about \SI{\UVOHpHsevenG}{V} vs SHE. The $V$--OH/$V\mathrm{H}_4$ boundary lies just above the $\mathrm{H_2}$ line, so full passivation needs roughly the hydrogen-evolution potential at pH~0 (the margin of \SI{0.02}{eV} is within our error) and a cathodic bias of about \SI{\UVHfourG}{V} at pH~7. Near-neutral water thus leaves the vacancy as the closed-shell oxygen A center. The hydroxyl complex, which keeps one dangling bond and is a paramagnetic center with a gap state, forms in acid or under mild cathodic bias, and the inert $V\mathrm{H}_4$ only under stronger cathodic bias. The vibrational terms do not change the order of the three regimes or the absence of the bare vacancy, but they move each boundary by several pH units. The $V$--OH window is the least certain part of the map. We did not compute a second hydrogen on $V$--OH; if it binds about as strongly as the later hydrogens at the bare vacancy, it would narrow this window or remove it.

\section{Discussion}

In silicon carbide the two sublattice vacancies bind hydrogen through different bonds. Removing a silicon atom leaves four carbon neighbors, so \VSi{} is hydrogenated through C--H bonds, while removing a carbon atom leaves four silicon neighbors. The ordering of their formation energies and the $D_{2d}$ distortion of \VC{} agree with earlier calculations~\cite{Zywietz1999}. \VC{} has the lower formation energy (\SI{\EfVCzSirich}{} vs \SI{\EfVSizSirich}{eV} under Si-rich and \SI{\EfVCzCrich}{} vs \SI{\EfVSizCrich}{eV} under C-rich conditions). The neutral and negative carbon vacancies undergo a $D_{2d}$ pairing distortion that brings silicon neighbors together to \SI{\nnminVCz}{\angstrom} and lowers the energy by \SI{\EJTVCz}{eV} relative to the symmetric structure. The neutral silicon vacancy relaxes outward, and its four carbon dangling bonds form a broken-symmetry singlet only \SI{\dEspinVSiz}{eV} below a ferromagnetic solution with a fractional moment of \SI{1.49}{\mu_B}.

Hydrogen behaves very differently at the two sites (Figs.~\ref{fig:structures}e,f and \ref{fig:model}a). At \VSi{} it forms an ordinary C--H bond (\SI{\XHVSiHz}{\angstrom}) and is bound by \SI{\EbVSi}{eV} relative to $\tfrac12\mathrm{H_2}$, more than four times as strongly as at the vacancy in silicon (\SI{\EbVH}{eV}). The resulting \VSi{}H is a doublet, with the spin on the remaining carbon dangling bonds. At \VC{} hydrogen does not form a terminal Si--H bond. We relaxed this complex from two different starting points, one with a Si--H cap and one pre-relaxed with a universal interatomic potential, and both ended in the same structure, in which the hydrogen bridges two silicon neighbors at \SI{\XHVCHz}{\angstrom} each and the other two silicon atoms pair. The binding energy of this three-center bond is only \SI{\EbVC}{eV}; the hydrogen is in fact unbound with respect to $\mathrm{H_2}$. Binding a hydrogen at \VC{} needs $\dmuH>\SI{\dmuVCH}{eV}$ before the vibrational correction, a hydrogen activity above that of $\mathrm{H_2}$ gas; by Eq.~(\ref{eq:che}) it is reached only under cathodic bias, below about \SI{-0.13}{V} vs SHE at pH~0. \VSi{} binds its first hydrogen already for $\dmuH>\SI{\dmuVSiH}{eV}$, far below any aqueous condition. These statements concern neutral cells and the first hydrogen only; the charged states of \VC{}H, oxygen and hydroxyl at \VC{}, and further hydrogenation of \VSi{} were not computed. In silicon the vibrational terms raise the formation energy by about \SI{0.2}{eV} per bound H, and a correction of similar sign would make the first hydrogen at \VC{} still less favorable. The first hydrogen need not decide the outcome, however: at the silicon vacancy hydrogen binds cooperatively, and hybrid-functional calculations for 4H-SiC find that \VC{} binds up to four hydrogens, which removes its gap levels~\cite{Cai2021}.

This asymmetry matters for devices because the carbon vacancy is responsible for the $Z_{1/2}$ center, the deep level that limits carrier lifetime in 4H-SiC power devices. Combined DLTS and EPR studies have assigned it to a negative-$U$ carbon-vacancy transition near $E_c-\SI{0.65}{eV}$, with the related $\mathrm{EH}_{6/7}$ levels deeper in the gap~\cite{Hemmingsson1997,Hemmingsson1998,Son2012,Klein2009,Kimoto2015,Coutinho2017}. A single hydrogen does little to this center: it binds only through the weak bridging configuration. Whether hydration deactivates the carbon vacancy therefore depends on the multiply hydrogenated complexes, which hybrid-functional calculations find to be stable and electrically inactive~\cite{Cai2021}, and on oxygen and hydroxyl at \VC{}, which we have not computed. Hydrogen at the SiC vacancies has also been studied in earlier supercell calculations~\cite{Aradi2001}. Measuring the $Z_{1/2}$ DLTS signal before and after a controlled aqueous or humid treatment would settle the question directly; in device processing $Z_{1/2}$ is normally removed by carbon-interstitial injection rather than by hydrogenation~\cite{Kimoto2015}.

Where a direct comparison is possible, the calculations agree with experiment and earlier theory: the host lattice constants, the bridging oxygen of the vacancy--oxygen A center, the Si--H stretch band of $V\mathrm{H}_{1\text{--}4}$~\cite{BechNielsen1996}, and, for SiC, the ordering of the vacancy formation energies and the $D_{2d}$ pairing of the carbon vacancy~\cite{Zywietz1999}. The main discrepancies concern the charged states. With our sampling the silicon vacancy does not distort, even with \SI{0.01}{eV} smearing, so the negative-$U$ ordering of its levels~\cite{Watkins1980,Anderson1975} is not recovered, and with the gap roughly halved, the acceptor levels of the vacancy and its hydrides end up at or above the PBE conduction-band edge. Neither problem changes the relative energies that determine the stability maps. A screened hybrid functional~\cite{Heyd2003,Deak2010} would be needed to place the acceptor levels correctly and recover the negative-$U$ ordering.

The same band-gap error, about a factor of two in both hosts, makes the absolute transition levels uncertain by a few tenths of an electronvolt, and the associated delocalization error likely contributes to the missing spin localization in $V\mathrm{H}$. Binding energies, disproportionation energies, and vibrational frequencies are ground-state, local quantities and much less sensitive, as the agreement with the measured LVMs suggests. The 64-atom cell with a shifted $2\times2\times2$ mesh is a compromise: compared with converged calculations it overestimates the neutral vacancy formation energy by about \SI{0.3}{eV} and misses the $D_{2d}$ relaxation~\cite{Probert2003,Corsetti2011}. Formation energies of the bare vacancy carry this error directly, whereas binding and reaction energies between complexes should be less affected. For a compact defect in a high-dielectric host the Makov--Payne plus alignment correction reproduces the FNV result to similar accuracy~\cite{Komsa2012}. The doubly charged state is the least converged case. The model is also purely thermodynamic. It gives the equilibrium decoration and charge state, but not the kinetics of hydrogen and hydroxyl in-diffusion, which decide whether equilibrium is reached during processing and explain why metastable $V\mathrm{H}_{1\text{--}3}$ complexes are seen after hydrogen implantation. Finally, we used 3C-SiC as the simplest cubic model of SiC. The inequivalent hexagonal and cubic sites of 4H-SiC shift the levels by a few tenths of an electronvolt~\cite{Coutinho2017}, but they should not change the C--H versus Si--H--Si asymmetry in hydrogen binding.

\section{Conclusions}

We have built and parameterized a grand-canonical model for the hydration of vacancies, in which the Fermi level and a water reservoir together set the charge state and the hydrogen/oxygen decoration of a vacancy. It combines charged-defect thermodynamics~\cite{Freysoldt2014}, the computational hydrogen electrode~\cite{Norskov2004}, and first-principles aqueous stability analysis~\cite{Persson2012,Todorova2014} in a single stability map, on which a defect can be followed from a bare, electrically active charge state to a decorated, passivated complex as the aqueous environment changes.

In 64-atom cells, DFT reproduces the host lattice constants to better than \SI{1}{\percent}, the bridging oxygen of the vacancy--oxygen A center, the $D_{2d}$ reconstruction of the SiC carbon vacancy, and, without scaling, the infrared Si--H stretch band of $V\mathrm{H}_{1\text{--}4}$ to within \SI{\LVMmaxpct}{\percent}. Hydrogen binds cooperatively to the silicon vacancy, so the partially hydrogenated complexes are never equilibrium species and the vacancy goes directly from bare to fully passivated. In contact with water the bare vacancy is never the lowest-energy state. With the vibrational free energy of the bound hydrogen and oxygen included, near-neutral water leaves it as the oxygen A center; among the complexes computed, the paramagnetic $V$--OH forms below about pH~\pHVOHG{} at zero bias; and full passivation to $V\mathrm{H}_4$ needs a cathodic bias. In equilibrium with a native oxide, which lowers the oxygen chemical potential by about \SI{2}{eV}, the oxygen complexes would lose this advantage. In PBE the charge state hardly enters, because the acceptor levels lie at or near the conduction-band edge; resolving the coupling between Fermi level and hydration will need a hybrid functional. In SiC, the first hydrogen binds strongly to the silicon vacancy through a C--H bond, whereas at the carbon vacancy, the lifetime-limiting $Z_{1/2}$ center in 4H-SiC devices, it forms only a weak three-center Si--H--Si bond and is unbound with respect to $\mathrm{H_2}$. Whether further hydrogens, as hybrid-functional calculations suggest~\cite{Cai2021}, or oxygen deactivate the carbon vacancy is the next question for this framework, and DLTS before and after a controlled aqueous treatment could settle it experimentally. The result also bears on SiC quantum technologies, where silicon-vacancy and divacancy spin qubits~\cite{Koehl2011} are exposed to hydrogen and hydroxyl during aqueous processing.

The same approach can be applied to other materials in contact with water, such as oxides, nitrides, halide perovskites, and two-dimensional materials, wherever charge state and hydration together control corrosion, photoelectrochemical stability, or dopant passivation.

\begin{acknowledgments}
The author thanks the Department of Chemical \& Biomolecular Engineering at the University of Delaware and the Delaware Energy Institute for financial support, the University of Delaware High Performance Computing facility for computational resources, and colleagues at the Delaware Energy Institute for helpful discussions.
\end{acknowledgments}

\section*{Data Availability}
The total energies, finite-size corrections, magnetic moments, local geometries, transition levels, and vibrational frequencies behind every figure and table, together with the relaxed structures (XYZ), the raw GPAW outputs, and the input and analysis scripts, are available from the author on reasonable request.

\bibliographystyle{apsrev4-2}
\bibliography{refs}

\end{document}